\documentclass{appolb}
\usepackage{epsfig}
\def\nostrocostrutto#1\over#2{\mathrel{\mathop{\kern 0pt \rlap
  {\raise.2ex\hbox{$#1$}}}
  \lower.9ex\hbox{\kern-.190em $#2$}}}
\def\gsim{\nostrocostrutto > \over \sim}   %greater or around...

\def\Journal#1#2#3#4{{#1} {\bf #2}, #3 (#4)}
\def\CPC{{\em Comp. Phys. Comm.} }
\def\EPJC{{\em Eur. Phys. J.} C}  
\def\IJMPA{{\em Int. J. Mod. Phys.} A}

\def\MPLA{{\em Mod. Phys. Lett.} A}

\def\PLB{{\em Phys. Lett.} B}

\begin{document}
% \eqsec  % uncomment this line to get equations numbered by (sec.num)
\title{Oscillating Hadron and Jet Multiplicity Moments
\thanks{Presented at the XXXIII Int. Symp. on Multiparticle
Dynamics,
Krakow, Poland, 5-11 September 2003,  based on work with M.A. Buican and C.
F\"orster \cite{bfo}.}%
% you can use '\\' to break lines
}
\author{Wolfgang Ochs
\address{Max-Planck-Institut f\"ur Physik (Werner-Heisenberg-Institut),
F\"ohringer Ring 6, D-80805 M\"unchen, Germany}
}
\maketitle
\begin{abstract}
Recently, the moments of multiplicity distributions in $e^+e^-$ annihilation
and the ratios $H_q$ (cumulant over factorial moments $K_q/F_q$) have been
determined both for the hadronic final state and for jets at variable
resolution. These ratios show an oscillatory behaviour as function of $q$
with strong dependence of the amplitude and length of oscillation on the jet
resolution parameter $y_{cut}$. The recent explanation of this phenomenon
based on perturbative QCD calculations is discussed. 
\end{abstract}
%\PACS{PACS numbers come here}
  
\section{Introduction}
A characteristic feature of high energy collisions is the multiparticle
production. It is generally understood as following from a primary hard
interaction (here $e^+e^-\to q\bar q$) with subsequent evolution of a parton
cascade according to the rules of perturbative QCD 
from the initial scale $Q$ down to a low
energy scale $Q_0$ where colour confinement leads to the formation of hadrons.
The parton evolution is dominated by gluon bremsstrahlung with its
collinear and soft singularities which causes the jet structure of the 
partonic final state also characteristic of the final state of hadrons.
The transition into hadrons is described by non-perturbative models.
In a particularly simple ansatz for hadronization one assumes
that observables for the multiparticle final state are proportional to the
corresponding quantities for partons, an idea which has been proposed
originally for single inclusive energy spectra and is called 
``Local Parton Hadron Duality'' (LPHD). Subsequently, 
this idea has been applied to a rather wide range of phenomena from inclusive 
multiparticle correlations to quasi-exclusive processes (for reviews, see
\cite{dkmt2,ko}). The agreement with data generally increases with the accuracy of
the calculations. 

Here we report on a recent application \cite{bfo} of these duality
ideas to the moments of multiplicity distributions which correspond to
integrals over multiparticle correlation functions. Experimental results
on these quantities have been presented 
by the L3 collaboration \cite{L3osc,mangeol}.

\section{Multiplicity moments}
We consider here the unnormalized and normalized factorial moments $f_q$
and $F_q$
\begin{equation}
f_q=\sum_{n=0}^\infty n(n-1)\ldots (n-q+1)P_n, 
   \quad  F_q=f_q/N^q, \quad  N\equiv f_1
\label{fmom}
\end{equation}
with multiplicity distribution $P_n$ and mean multiplicity $N$.
Furthermore, one introduces the cumulant moments
$k_q$ and $K_q$ which are used to measure the genuine correlations without  
 uncorrelated background in a multiparticle sample
\begin{equation}
k_q=f_q-\sum_{i=1}^{q-1} {q-1 \choose i} k_{q-i} f_i, \qquad K_q=k_q/N^q,
\label{kmom}
\end{equation}
in particular $K_2=F_2-1,\ K_3=F_3-3F_2+2$; for a Poisson distribution
$K_1=1,\ K_q=0$ for $q>1$.
Of special interest are
the ratios 
\begin{equation}
H_q=K_q/F_q
\end{equation}
as there is a prediction for high energies of
an oscillatory behaviour \cite{dreminosc}
with the first minimum near $q\approx5$ at LEP energies.
Such a minimum has been observed indeed in
$e^+e^-$ annihilations at SLC \cite{slacosc} and  at LEP
\cite{L3osc,mangeol} but the magnitudes of moments are found much smaller
than predicted \cite{lo}.

\section{QCD predictions}
Predictions in \cite{bfo} are obtained by solving numerically the evolution equations for the
moments which are derived from the evolution equations of the single jet
generating function (review \cite{dkmt2}) together with the initial 
condition at
threshold at fixed transverse momentum cut-off $Q_c$. This cut-off
characterizes at the same time the hadronization scale $Q_c=Q_0\gsim
\Lambda_{QCD}$ in the description of the hadronic final state and also
the jet resolution (parameter $y_{cut}=(Q_c/Q)^2,\ Q=2E$ at jet energy $E$) 
for the jet final state obtained with a jet algorithm based on a
transverse momentum cut-off (``Durham algorithm''). 
Different approximations are considered.

In the  Double Logarithmic Approximation (DLA) one takes into account the
most singular bremsstrahlung singularities. This corresponds to the asymptotic
solution and some results can be given in analytic form, for example,
$H_q\to1/q^2$. These asymptotic results are quickly approached above threshold
energy $E=Q_0$ where there is only one particle in the
jet. Then, $F_q=0$ for $q>1$ and $K_q$ oscillates
\begin{equation}
K_q=(-1)^{q-1}(q-1)!\quad {\rm for} \ Y_c=\ln(E/Q_c)=0   \label{kmom0}
\end{equation}
and consequently also $H_q$ does.
\begin{figure}[t!]%[hbt]
\begin{center}
\includegraphics[angle=-90,width=12.5cm]{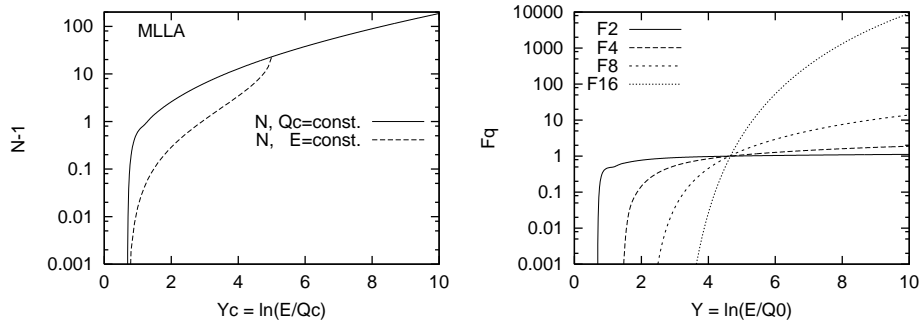}
\end{center}
\vspace{-4.5cm}
\caption{
Multiplicity $N$ in MLLA in single jet
 vs energy for fixed $Q_c=Q_0$ 
(representing hadrons) and for fixed energy $Y_0=\ln(E/Q_0)=5$
 but variable $Q_c$ representing
jets at variable resolution $Q_c$ and 
factorial moments $F_q$ vs. energy $Y$. 
}
\label{hqmlla-nf}
\end{figure}    

The Modified Logarithmic Approximation (MLLA) takes into account certain
non-leading logarithmic terms and gives a more realistic description at finite
energies. In Fig. \ref{hqmlla-nf} the energy dependence of the parton 
multiplicity (cut-off $Q_0$ for hadrons) is seen above the multiplicity for
jets at the same ratio $E/Q_c$ 
because the transverse momentum is in the argument
of the coupling constant $\alpha_s$ and $Q_c>Q_0$ (see also \cite{lo}). 
Contrary to the DLA
the threshold for higher moments $F_q$ is shifted towards higher energies.
Remarkably, there is a Poissonian transition point with  $F_q=1$ 
for all moments. Below this point the $H_q$ ratios
show rapid oscillations as in (\ref{kmom0}) whereas above there is a
transition to  oscillations of length increasing with energy. 

\begin{figure}[t!]%[hbt]
\begin{center}
%\mbox{\epsfig{file=p-001070-Model.EPS,width=8.cm,bbllx=1.cm,bblly=5.5cm,%
%bburx=19.cm,bbury=27.5cm}}
%\vspace{-0.5cm}
\mbox{\epsfig{file=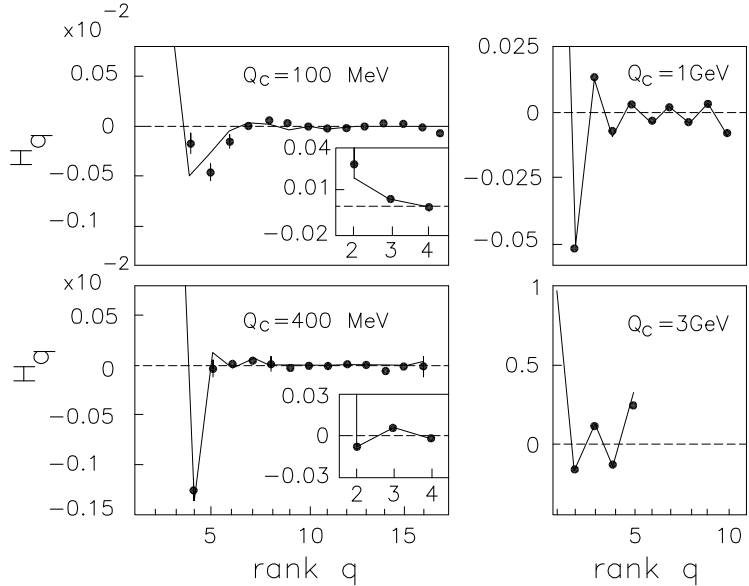,width=10cm,clip=,bbllx=2.cm,bblly=11.7cm,%
bburx=20.cm,bbury=26.5cm}}
\end{center}
\vspace{-0.5cm}
\caption{
Ratios $H_q$ of multiplicity moments for jets at different cut-off $Q_c$ 
as obtained by L3 Collaboration \protect\cite{L3osc} compared with our 
MC results (ARIADNE-D).
}
\label{L3oscillations}
\end{figure}

Finally, a comparison of data is performed with a Monte Carlo calculation
of the parton cascade where we use ARIADNE-D as in \cite{ARIADNE} but with
refit parameters $Q_0$ and $\Lambda_{QCD}$ 
to simulate the hadronic final state directly. Here energy
momentum conservation is fully taken into account as well as the
$O(\alpha_s)$ matrix element for $e^+e^-\to q\bar q g$. The results are
similar to those from the MLLA 
evolution equation. In particular,
the data on jet and hadron multiplicities with the splitting as in Fig.
\ref{hqmlla-nf} are rather well described and also the variation of the
oscillation pattern of $H_q$ with jet resolution $y_{cut}=(Q_c/Q)^2$, see
Fig. \ref{L3oscillations}, in further support of the duality approach.


\begin{thebibliography}{99}
\bibitem{bfo}
M.A. Buican, C. F\"orster and W. Ochs, \Journal{\EPJC}{31}{57}{2003}

\bibitem{dkmt2}
Yu.L. Dokshitzer, V.A. Khoze, A.H.
Mueller and S.I. Troyan, \lq\lq Basics of Perturbative
QCD", ed. J. Tran Thanh Van, Editions Fronti\'{e}res,    
Gif-sur-Yvette, 1991.

\bibitem{ko} 
V.A.  Khoze and W. Ochs,
\Journal{\IJMPA}{12}{2949}{1997};
%Int. J. Mod. Phys. {\bf A12} (1997) 2949;
V.A. Khoze, W. Ochs and J. Wosiek, in ``Handbook of QCD'', ed. M.A. Shifman
(World Scientific, Singapore, 2001), arXiv:hep-ph/0009298.

\bibitem{L3osc}
L3 collaboration, P.Achard et al., \Journal{\PLB}{577}{109}{2003};
%Phys. Lett. B577, 109 (2003);
%"Measurement of the Charged Particle Multiplicity
%and Inclusive Momentum Distributions in Z Decays at LEP", 
L3 Note 2676,
submitted to Int. Europhysics Conf. on HEP (paper 512), Budapest, 2001,
{\tt http://l3.web.cern.ch/l3/conferences/Budapest2001/papers/\\%
note\_2676/note\_2676.ps.gz} (June 25, 2001).
 

\bibitem{mangeol}   
D.J. Mangeol, 
%``Correlations in the Charged -Particle Multiplicity Distribution'', 
thesis Univ. of Nijmegen, Jan. 2002,
arXiv:hep-ex/0110029. 

\bibitem{dreminosc}
I.M. Dremin,  \Journal{\PLB}{313}{209}{1993};
     % Phys. Lett. B313 (1993) 209. 
I.M. Dremin and V.A. Nechitailo,   \Journal{\MPLA}{9}{1471}{1994}.
 %Mod. Phys.\Lett. {\bf A9} (1994) 1471.

\bibitem{slacosc}
SLD collaboration, K. Abe et al.,  \Journal{\PLB}{371}{149}{1996}.
% Phys. Lett. B371 (1996) 149.

\bibitem{lo}
S. Lupia and W. Ochs,   \Journal{\PLB}{418}{214}{1998}.
                       %Phys. Lett. {\bf B418} (1998) 214;
%
%\bibitem{lu}
S. Lupia,  \Journal{\PLB}{439}{150}{1988}.  


\bibitem{ARIADNE}   % 34  
  L.\ L\"onnblad,
                                 \Journal{\CPC}{71}{15}{1992}.
  % {\it Comp. Phys.\ Comm.} {\bf 71} (1992) 15. %ll


\end{thebibliography}
\end{document}